\documentclass[12pt]{article}
  \usepackage{amsfonts}
\newcommand{\ba}{\begin{array}}
\newcommand{\ea}{\end{array}}
\newcommand{\be}{\begin{equation}}
\newcommand{\ee}{\end{equation}}
\newcommand{\bea}{\begin{eqnarray}}
\newcommand{\eea}{\end{eqnarray}}
\font\sqi=cmssq8
\def\DR{\rm I\kern-1.45pt\rm R}
\def\DC{\kern2pt {\hbox{\sqi I}}\kern-4.2pt\rm C}
\def\DH{\rm I\kern-1.5pt\rm H\kern-1.5pt\rm I}

\begin{document}
\begin{center}
{\bf \large Anisotropic    Higgs oscillator }\\[4mm]

{\large Armen Nersessian$^{1,2}$ and  Vahagn Yeghikyan$^{1}$}
\end{center}

{\sl $\;^1$  Yerevan State University,  Alex Manoogian St. 1, Yerevan, 375025

 $\;^2$   Artsakh State University, Mkhitar Gosh St. 3,
 Stepanakert

 $\;^{\;}$  Yerevan Physics Institute,  Alikhanian Brothers St. 2,
 Yerevan
 \begin{center} 
  ARMENIA
  \end{center}}
    \begin{abstract}
\noindent We present  the higher-dimensional generalization of anisotropic (pseudo)spherical  oscillator
suggested recently in  [ArXiv 0710.5001], and the related spherical and pseudosperical generalizations
(MICZ-)Kepler like systems with Stark term and $\cos\theta$ potential.
We write down the potentials and hidden symmetry generators of these systems, in terms of ambient Euclidean space.
\end{abstract}

\noindent The oscillator  plays a  distinguished role among
mechanical systems with hidden symmetries. Due to enormous number
of   hidden symmetries it preserves its integrability property
after numerous deformations.  Most known among  them are  the anisotropic oscillator and
the isotropic oscillator in the presence of constant
magnetic  field.
The are also less known, nonlinear   deformations of the oscillator,  preserving
the integrability property, e.g. the $2p$-dimensional oscillator deformed by the potential
\be U_{nlin}=-2 \varepsilon_{el}\sum_{i=1}^{p}
x^4_i-x^4_{i+p}.\label{nlin}\ee
An interesting peculiarity of the oscillator is its relation with the Kepler system.
 Namely,
$(p+1)-$dimensional Kepler system  can be obtained by the
appropriate reduction procedures from the $2p-$dimensional
oscillator,  for $p=1,2,4$ (for the review see, e.g.
\cite{terant}). These procedures, which  are known under the
names Levi-Civita (or Bohlin) \cite{bohlin}, Kustaanheimo-Stiefel \cite{ks}
and Hurwitz \cite{h} transformations,  imply
 the reduction of the oscillator  by the  action of $Z_2$,
$U(1)$, $SU(2)$ group, respectively, and yield the Kepler-like
systems with hidden symmetry, which are  specified by the presence
of monopoles \cite{ntt,mic,su2}.
 The second system (with $U(1)$ (Dirac) monopole) is most
known and the most important among them. It was invented independently by
Zwanziger and by McIntosh and Cisneros \cite{micz}. Presently it is
called  MICZ-Kepler system.
Let us notice, that the (pseudo)spherical analog of the two-center Kepler system \cite{otchik},
as well as  the two-center MICZ-Kepler system
\cite{kno}, are also integrable systems. The (pseudo)spherical two-center MICZ-Kepler system \cite{no}
 seems also to be  integrable system.

After similar reductions the deformed oscillators result in the integrable deformations of the Coulomb systems.
For example, the  $2p$-dimensional oscillator
 with additional anisotropic term
\be U_{AI}=\frac{\Delta\omega^2}{2}\sum_{i=1}^{p}(x^2_i-x^2_{i+p})
\label{delta}\ee
 results in the $(p+1)-$dimensional  Kepler system with potential
\be
 V_{cos}= \frac{\Delta\omega^2}{4}\cos\theta=\frac{\Delta\omega^2}{4}\frac{x_{p+1}}{|{\bf x}|},
  \label{cos}\ee
while the oscillator deformed by the potential  (\ref{nlin})
results in the $(p+1)-$dimensional
Kepler system with linear potential, which is the
textbook example of the deformed Kepler system admitting the
separation of variables
 in parabolic  coordinates.

The oscillator  admits the  generalizations to a $d-$di\-men\-si\-onal
sphere and a two-sheet hyperboloid (pseudosphere) which also possesses the hidden nonlinear symmetry.
This  generalization was suggested by  Higgs \cite{sphere}. Presently it is referred  as Higgs oscillator.
 The Higgs oscillator is defined  by the potential
\begin{equation}
U_{osc}=\frac{\omega^2 R_0^2}{2}\frac{{\bf x}^2}{{x}^2_{0}},
\label{v}\end{equation}
where ${\bf x}, x_{0}$ are the Cartesian  coordinates of the ambient
(pseudo)Euclidean space $\DR^{d+1}$($\DR^{d.1}$):
$\epsilon{\bf x}^2+ x^2_{0}=R_0^2$, $\epsilon=\pm 1$.
 The $\epsilon=+1$ corresponds
to the sphere and  $\epsilon=-1$ corresponds to  the pseudosphere.
This system has the same number of hidden symmetry generators, like a planar one,
which are given by the expression
\begin{equation}
{ A}_{\alpha\beta}=\frac{{J}_\alpha{ J}_\beta}{2 R_0^2} +
\omega^2 R_0^2\frac{x^\alpha{x}^\beta}{2x^2_0},
\label{Ic}\end{equation}
where $J_\alpha$ are the translation generators along $x_\alpha$ direction.

 Consequently, the properties of the Higgs oscillator are quite similar to the properties of conventional oscillator.
For example, Higgs oscillator has closed classical trajectories and degenerated energy spectrum,
it admits the separation of variables in a few coordinate systems,  etc.
Moreover,
the $2p$-dimensional pseudospherical Higgs oscillator  can be related (completely similar to the planar case) with
$(p+1)$-dimensional  pseudospherical (MICZ-)Kepler system, and
  $2p$-dimensional spherical Higgs oscillator  can be related  with
$(p+1)$-dimensional  pseudospherical (MICZ-)Kepler system
\cite{np}.
 After appropriate ``Wick rotation"  (similar to the
 one performed in \cite{kmp}) of the MICZ-Kepler system on hyperboloid, one can obtain the
MICZ-Kepler system on the sphere constructed in \cite{kurochkin}.
Let us remind that (pseudo)spherical Kepler system, which possess the hidden symmetries, has been suggested by
Schroedinger \cite{sphere1}.
It is defined by the potential
\be
 V_{Kepler}=
-\frac{\gamma}{R_0}\frac{x_{0}}{|{\bf x}|}.
\ee
Hence, one can expect that the Higgs oscillator could  have the anisotropic integrable generalization,
similar to the conventional one.
This generalization, which was found in our  recent paper  \cite{anosc},
is defined by the potential
\be \frac{\Delta\omega^2}{2}
 {\bf x} {\widehat T} {\bf x},\label{sadelta} \ee
 where $({\widehat T}$ is $d\times d)$ symmetric matrix which obeys the conditions
 \be
  {\widehat T}^2={\widehat {\rm Id}},\qquad {\widehat T} \neq {\widehat {\rm Id}}.
\label{T}\ee
The hidden symmetry generators of  this anisotropic oscillator read
 \begin{equation}
A={A}_T + \frac{\Delta\omega^2}{2}{{\bf x}{\widehat T}{\bf x}}, \quad A_T \equiv T_{\alpha\beta} A_{\alpha\beta}.
 \label{IA}\end{equation}
The (pseudo)spherical analog of the oscillator with additional potential  (\ref{nlin}) was found  in \cite{anosc} as well.
It is defined by the Higgs oscillator Hamiltonian with additional potential term
\be
U_{nlin}=
\varepsilon_{el}R^2_0\frac{(R_0^2+x^2_0)}{x_0^4}({\bf x}{\bf x}) ({\bf
x} {\widehat T} {\bf x}). \label{inhasph}\ee
The hidden symmetry generator of latter  system is given by the expression
 \begin{equation}
A={A}_T
+ 4\varepsilon_{el}\left(
 \frac{R^2_0{\bf x}^4}{x^2_0}+
\frac{R^4_0({\bf x}\widehat T{\bf x})^2}{x^4_0}\right) . \label{IA1}
\end{equation}

With these potentials at hand one can get the integrable pseudospherical analog of Kepler system with linear  and
$\cos\theta$ potentials. For this purpose, we must  perform the Kustaaheimo-Stiefel transformation of the
four-dimensional anisotropic Higgs oscillator, and obtain the three-dimensional (MICZ-)Kepler
system with the additional potential given by the expression \cite{anosc},
 \be V_{AI}=\frac{\Delta
\omega^2}{2}\left(\frac{x_3}{|{\bf x}|}+\epsilon
\frac{x_0x_3}{R^2_0}\right)+\varepsilon_{el}\frac{x_0 x_3}{R_0}
\label{sstark}\ee
  We must perform
 its ``Wick rotation"  and take its  real part, to get the spherical counterpart of the obtained system.
 In terms of ambient space coordinates the additional potential of the resulting
 spherical
 system will be given by the same expression, as the one of the pseudospherical system (\ref{sstark}).

The  hidden symmetry generator  of the obtained system is defined by the expression
\be A=
A_3+ \left(2 \varepsilon_{el}+\frac{  \Delta \omega^2}{|{\bf x}|} \right)
\left[
{\bf x}^2-x^2_3\right] \ee
where $A_3$ is the third component of the (pseudo)spherical Runge-Lenz vector.
That is why we  define the   (pseudo)spherical analog  of the linear potential
by the expression
\be
V_{Stark}= \varepsilon_{el}
\frac{x_0}{R_0} x_{p+1}.
\ee
Besides, we can define the following (pseudo)spherical analog of the potential (\ref{cos})
 \be V_{\cos}=\frac{\Delta
\omega^2}{2}\frac{x_{p+1}}{|{\bf x}|}.\label{arp}\ee

So, we presented  the integrable (pseudo)spherical generalization  of anisotropic oscillator which can be viewed as
a deformation of the well-known
Higgs oscillator. By the use of this system, we constructed the integrable (pseudo)spherical analog of the MICZ-Kepler system
with the linear and $\cos\theta$ potentials.\\

{\large Acknowledgments.} We thank  Levon Mardoyan, Vadim Ohanyan  and
George Pogosyan for useful comments. A.N. is grateful to the Organizers of International
Workshop on {\sl Supersymmetries and Quantum Symmetries} for the kind invitation at  Workshop
 and for the hospitality in Dubna.
The work is supported by the grant of Artsakh Ministry of Education and Science,
 by the  NFSAT-CRDF  grants ARP1-3228-YE-04 and UC 06/07,
and by the
  INTAS-05-7928  grant.


\begin{thebibliography}{99}

\bibitem{terant}V.~M.~Ter-Antonyan, {\sl  Dyon-Oscillator duality},
[quant-ph/0003106]




 L.~G.~Mardoyan {\sl et al}
 {\em  Quantum systems with hidden symmetry}, MAIK Publ., Moscow, 2006




\bibitem{bohlin}T. Levi-Civita, Opere Mathematiche,{\bf 2}(1906), 411

K.~Bohlin, Bull.~Astr., {\bf 28} (1911), 144

\bibitem{ks}P.~Kustaanheimo,~E.~Stiefel,~J.~Reine Angew Math., {\bf 218} (1965), 204
\bibitem{h}A.~Hurwitz,~Mathematische~Werke,~Band~II,~641
({\sl Birkh\"auser},~Basel,~1933)

L.~S.~Davtyan {\it et al}, J.~Phys.~{\bf A20}~(1987),~6121;

D.~Lambert, M.~Kibler, J.~Phys.~{\bf A21}~(1988),~307
\bibitem{ntt}A.~Nersessian,~V.~M.~Ter-Antonyan,~M.~Tsulaia, Mod.~Phys.~Lett.
{\bf A11}~1605~(1996);
\bibitem{mic}T.~Iwai,~Y.~Uwano, J.~Math.~Phys.~{\bf 27}~(1986),~1523

             I.M. Mladenov and V.V. Tsanov, 
             {J. Phys.}, {\bf A20}, 5865 (1987);  {\bf A32}, 3779 (1999).
\bibitem{su2}T.~Iwai,~J.~Geom.~Phys.~{\bf 7}~(1990),~507;

L.~G.~Mardoyan,~A.~N.~Sissakian,~V.~M.~Ter-Antonyan, Phys.~Atom.~Nucl. {\bf 61}~(1998),~1746

      \bibitem{micz}   D.~Zwanziger,{Phys.~Rev.} {\bf 176},~1480 ~(1968)

      H. McIntosh and A. Cisneros, {J. Math. Phys.} {\bf 11}, 896    (1970).

\bibitem{otchik}V.~S.~Otchik, Doklady Akademii Nauk BSSR, {\bf 35}, No. 5, 420 (1991)(in Russian)
  \bibitem{kno}S.~Krivonos, A.~Nersessian and  V.~Ohanyan
 {\em Phys.\ Rev.\/} {\bf D75}, 085002 (2007)
\bibitem{no}A.~Nersessian and V.~Ohanyan,
 [arXiv:0705.0727],
 to appear in {\sl Theor. Math. Phys}
\bibitem{sphere}P.~W.~Higgs,~J.~Phys.~A:~Math.~Gen.~{\bf 12}~309~(1979)

H.~I.~Leemon,~J.~Phys.~A:~Math.~Gen.~{\bf 12}~489~(1979)


\bibitem{np} A.~Nersessian and G.~Pogosyan,
 {\em Phys.\ Rev.\/} {\bf A63}, 020103(R) (2001)

\bibitem{kmp}E.~G.~Kalnins,~W.~Miller~Jr.,~G.~S.~Pogosyan,
J.~Math.~Phys.{\bf 37} 2629(2000)
 \bibitem{kurochkin}
V.~V.~Gritsev, Yu.~A.~Kurochkin, V.~S.~Otchik,
{\em J.~Phys.\/} {\bf A33}, 4903-4910 (2000)

\bibitem{sphere1}E.~Schr\"odinger,~Proc.~Roy.~Irish~Soc.~{\bf 46}~9~(1941);~{\bf 46}~183
(1941);~{\bf 47}~53~(1941)

\bibitem{anosc}A.~Nersessian and V.~Yeghikyan,  [arXiv:0710.5001]
\end{thebibliography}
\end{document}